%% file: main.tex
\documentclass[10pt]{IEEEtran}

\usepackage{tikz}
\usepackage{amsmath}
\usepackage{xspace}
\usepackage{enumitem}
\usepackage{amsfonts}
\usepackage{graphicx}
\usepackage{tikz}
\usepackage{amsmath}
\usepackage{times}
\usepackage{subfigure}
\usepackage{caption} 
\usepackage{algorithm}
\usepackage{algorithmic}
\usepackage{tabularx}
\usepackage{xspace}
\usepackage{booktabs}
\usepackage{multirow}
\usepackage{enumitem}
\usepackage{makecell}
\setlist{nosep}
\usepackage{url}
\usepackage{orcidlink}

\hypersetup{
  colorlinks=false,
  linkbordercolor=white,
 urlbordercolor=white,
pdfborder={0 0 0}
}

\newcolumntype{C}[1]{>{\centering\arraybackslash}p{#1}}

\newcommand{\system}{{\textsc{\small{FHEON}}}\xspace}

\newcommand{\includeimg}[1]{\includegraphics[width=\linewidth]{#1}}

\begin{document}

\title{FHEON: A Configurable Framework for Developing Privacy-Preserving Neural Networks Using Homomorphic Encryption}

\author{
Nges Brian Njungle \orcidlink{0009-0006-3393-6851}, Eric Jahns \orcidlink{0009-0004-5511-7975}, and Michel A. Kinsy  \orcidlink{0000-0002-1432-6939} \\
STAM Center, Ira A. Fulton Schools of Engineering\\
Arizona State University, Tempe, AZ 85281, USA\\
Emails: nnjungle@asu.edu, jjahns@asu.edu, mkinsy@asu.edu
}

\maketitle

\input{sections/abstract}
\begin{IEEEkeywords}
Privacy-preserving Machine Learning, Neural Networks, Homomorphic Encryption, OpenFHE, FHEON
\end{IEEEkeywords}

\input{sections/introduction}

\input{sections/relatedworks}
\input{sections/background}

\input{sections/threat_model}
\input{sections/cnnlayers}

\input{sections/fhelayers}
\input{sections/experiment}
\input{sections/conclusion}

\bibliographystyle{plain}
\bibliography{paper}

\appendix

\input{sections/appendix/architectures}

\end{document}

%% file: sections/abstract.tex
\begin{abstract}
The widespread adoption of Machine Learning as a Service raises critical privacy and security concerns, particularly about data confidentiality and trust in both cloud providers and the machine learning models. 
Homomorphic Encryption (HE) has emerged as a promising solution to this problems, allowing computations on encrypted data without decryption. 
Despite its potential, existing approaches to integrate HE into neural networks are often limited to specific architectures, leaving a wide gap in providing a framework for easy development of HE-friendly privacy-preserving neural network models similar to what we have in the broader field of machine learning.

In this paper, we present \system, a configurable framework for developing privacy-preserving convolutional neural network (CNN) models for inference using HE. \system introduces optimized and configurable implementations of privacy-preserving CNN layers including convolutional layers, average pooling layers, ReLU activation functions, and fully connected layers. 
These layers are configured using parameters like input channels, output channels, kernel size, stride, and padding to support arbitrary CNN architectures.
We assess the performance of \system using several CNN architectures, including LeNet-5, VGG-11, VGG-16, ResNet-20, and ResNet-34. \system maintains encrypted-domain accuracies within $\pm1\%$ of their plaintext counterparts for ResNet-20 and LeNet-5 models. 
Notably, on a consumer-grade CPU, the models build on \system achieved $98.5\%$ accuracy with a latency of $13$ seconds on MNIST using LeNet-5, and $92.2\%$ accuracy with a latency of $403$ seconds on CIFAR-10 using ResNet-20. 
Additionally, \system operates within a practical memory budget requiring not more than $42.3$~GB for VGG-16. 
\end{abstract}

%% file: sections/introduction.tex
\section{Introduction}
\label{sec:intro}

Machine Learning as a Service (MLaaS) has emerged as a cornerstone for deploying machine learning (ML) models at scale, leveraging cloud platforms' computational power and storage capabilities \cite{9194237}. Model providers typically use cloud infrastructures to train models on diverse datasets and then commercialize the resulting models by offering them as services hosted on the cloud. However, the adoption of MLaaS introduces critical privacy and security challenges. When training datasets originate from multiple private sources, data contributors may hesitate to share their data due to trust concerns in the model owner or cloud provider. Similarly, in scenarios where pre-trained models are offered as services, the model's end users may be reluctant to submit sensitive data to the model provider or cloud infrastructure. 
In some cases, applications rely on sensitive data, such as private health records, and must strictly adhere to data protection regulations and laws, including HIPAA and GDPR \cite{mcgraw2021privacy} \cite{negri2024understanding}.

These challenges highlight a pressing need for mechanisms that protect both data and models throughout the MLaaS pipeline. 
Homomorphic Encryption (HE) has emerged as a promising solution by enabling computations to be performed directly on encrypted data, thereby offering strong privacy with cryptographic security guarantees rooted in hard mathematical assumptions and minimal communication overhead \cite{pulido2020survey, guardianml}. 
In the context of MLaaS, HE provides a natural defense against untrusted service providers, addressing the trust concerns of both data owners and end users.
Despite these advantages, the practical deployment of HE in MLaaS remains severely constrained by its high computational and memory costs. 
Convolutional neural networks (CNNs) pose a particularly unique challenge as their implementations require a combination of multiple different linear layers and non linear layers \cite{acar2018survey}.
These constraints make it difficult to directly adapt existing CNN implementation methodologies to HE, forcing researchers to come-up with new strategies and optimizations for the efficient implementation of CNNs under HE constraints.  
Prior efforts have proposed multiple optimizations but their designs are generally focused on a very narrow space such as a specific class of CNNs or even a single architecture (mainly the ResNet-20 architecture) \cite{rovida_cnn, Joon}.
Such solutions are not guaranteed to generalize across the diverse neural network classes and architectures  widely used in practice  and therefore limit applicability in real-world MLaaS pipelines.
What is missing is a unified, reusable, and efficient framework that provides general-purpose encrypted CNN building blocks with system-level optimizations. 
This framework will enable practitioners to easily develop privacy-preserving CNNs under HE constraints without reinventing or re-implementing all the core HE-friendly techniques.
By abstracting away the complexities of HE for all CNN layers,  such a system allows developers and researchers to concentrate on their model design as well as application specific goals, rather than reinventing low-level HE-friendly neural network optimizations(which are generally very difficult to re-implement and time consuming) needed by their work just like in the broader ML field.

To bridge this gap, we present \system, a robust, efficient, and configurable framework for building privacy-preserving convolutional neural networks (CNNs) for encrypted inference using the CKKS scheme \cite{ckks}. 
Unlike prior work, which often propose optimizations and supports a narrow set architectures or specific classes of CNNs, \system provides reusable and parameterized HE-friendly layers and a set of system-level optimizations that collectively make encrypted CNN inference development easy and feasible even on  consumer-grade hardware.
At the core of \system are highly optimized and configurable encrypted CNN building blocks like convolution, average pooling, ReLU activation, and fully connected layers. 
These building blocks are designed to support different parameters for common CNN layers such as striding, padding,  kernels, input channels, and output channels. 
We further introduce different methods for importing model weights  across heterogeneous training pipelines and discuss evaluation keys reuse strategies that significantly reduce both runtime and memory overhead for inferring on encrypted data.

To demonstrate the effectiveness of \system, we instantiate several widely-used CNN models, including LeNet-5, VGG-11, VGG-16, ResNet-20, and ResNet-34, and evaluate them on MNIST, CIFAR-10, and CIFAR-100. 
Our models preserve high accuracy (e.g., $98.5\%$ on MNIST with LeNet-5 and $92.2\%$ on CIFAR-10 with ResNet-20) while operating within practical resource limits. 
Notably, \system achieves the lowest reported memory footprint for encrypted ResNet-20 inference ($20.4$ GB) as well as latency (403 seconds).  \system also ensures that even deeper networks like VGG-16 remain feasible ($42.3$ GB), substantially lowering the barrier to real-world adoption of HE-friendly CNNs.

\noindent Concretely, our contributions in this work are:
\begin{itemize}
\item  We design and develop \system, an open-source framework that provides optimized, modular, parameterized, and HE-compatible CNN layers, enabling end-to-end encrypted inference across diverse neural network architectures and classes. 
Our HE-compatible layers support generic neural network configurations such as striding, padding, kernel sizes, and input and output channels.  
\item We design and implement multiple evaluation-key generation and re-use mechanisms, substantially lowering the memory footprint and computational overhead required for encrypted inference.
\item We build a flexible import path for model weights by allowing flexible weights processing for encrypted inference .
We also support multiple easy-to-use and configurable helper functions that facilitate the easy setup of the cryptographic context for HE-friendly models.
\item  We instantiate and benchmark multiple representative CNNs across multiple datasets, showing that \system consistently preserves accuracy while achieving the lowest reported memory usage for encrypted CNN inference on a consumer-grade CPUs.
\end{itemize}
\system is publicly available at 
\url{https://github.com/stamcenter/fheon}, with a comprehensive user 
documentation of modules and examples available at 
\url{https://fheon.pqcsecure.org}.

%% file: sections/relatedworks.tex
\section{Related Works}
\label{sec:related_works}

Privacy-preserving neural networks that leverage HE have made remarkable progress since the introduction of CryptoNets in 2016  \cite{gilad2016cryptonets}. These advances are generally classified into two generations based on the efficiency of computation and the depth of neural networks evaluated. 

The first generation CNN works adopted HE for use by applying two main optimizations. Firstly, they replaced the activation functions in networks with low-degree polynomials, such as the square function during training and inference. Based on the custom activation functions, they designed custom networks of limited depths for optimal performance and accuracy. Some of the works in this generation include: CryptoNets \cite{gilad2016cryptonets}, CryptoDL \cite{hesamifard2017cryptodl}, and Al Badawi et al. \cite{al2020towards}. 
While these simplifications reduced computational complexity, they limited the performance of models as well as scalability thus, these models were only evaluated on very simple tasks. 

The second generation of CNNs with HE introduced more sophisticated architectures with significant performance improvements. These works employed sophisticated techniques, such as bootstrapping, enabling computations of arbitrary-depth circuits, and processing privacy-centric, complex, and deep networks. During training, non-linear activation functions  are utilized, while the polynomial approximations of the functions are used during encrypted inference. 
Works in this generation generally adopt the CKKS or TFHE schemes. 
The TFHE scheme is known for its efficient bootstrapping carried out after every multiplication \cite{tfhe}. On the downside, TFHE is inherently slow in large-scale data processing compared to CKKS due to its lack of support for parallel data processing. 
Secondly, TFHE operates on bit-wise data, which contrasts with the floating-point arithmetic generally used in ML applications, thus it is not tractable for processing complex tasks and deep CNN architectures \cite{narumanchi2017performance}. 
Homomorphic CNN works that use TFHE rely on small models, consisting of only a few layers. These models generally show low accuracies and significant latencies on complex tasks compared to works based on CKKS. A prominent examples is Badawi at al. \cite{badawi2023}.

The CKKS scheme has emerged as the most compatible HE scheme for ML works as it is the only mainstream HE scheme that supports floating-point arithmetic and parallel data processing \cite{lee2022privacy}. These capabilities significantly accelerate computations while providing a possible direct mapping of functionalities with unencrypted ML work. In 2022, Lee et al. proposed an HE framework that achieved 91.31\% accuracy on the CIFAR-10 dataset in 2,271 seconds \cite{lee2022privacy}. Building on this, Kim et al. introduced a more efficient approach, achieving 92.04\% accuracy on CIFAR-10 with a single image inference time of 255 seconds \cite{kim2023optimized}. 
The works of Lee et al. and  Kim et al. require substantial memory, as Lee et al.’s framework requires over 512 GB of memory for computation, while Kim et al. reduce the memory requirement to about 100 GB. In 2024, Rovida et al. proposed an optimized implementation of Kim et al.’s work, achieving 91.53\% accuracy on the CIFAR-10 dataset in just 260 seconds while using only 15.1 GB of memory \cite{rovida_cnn}. It is worth nothing that these works were mostly based on ResNet-20 models, highlighting the limited CNN architectures evaluated by HE-friendly CNN works of today.

While the aforementioned works provide HE-based CNN models for inference, none proposed a configurable approach similar to those used in the broader field, where diverse architectures are developed from a common base. 
Instead, Zama tries to solve this problem by developing ConcreteML, a framework that simplifies the development of TFHE-friendly ML models through a Pytorch-like interface \cite{ConcreteML}. While basic ML workloads are very efficiently on ConcreteML, developing deep neural networks is very challenging, slow, and non-tractable as it relies on the TFHE scheme implementation in the Concrete Library, also supported by Zama. 
In contrast, our approach to HE-based CNN inference shifts the focus from optimizing architectures using CKKS to creating an efficient, scalable, and configurable framework that enables users to easily develop arbitrary CNN models on CKKS. 
\system provides configurable CNN layers with optimizations for efficient usage of resources, dynamic keys management, and weights loading using the CKKS scheme, thus ensuring the practicality of developing HE-friendly CNN inference models for diverse use cases. We further evaluate our work on a customer-graded CPU to show the compatibility of \system for use in everyday systems significantly outperforming related works in performance and resource usage. 

Other research works that focus on  HE-friendly CNNs have explored two primary areas: high-throughput model design and activation function optimizations. Approaches targeting high-throughput models typically employ batching techniques to amortize computation across multiple inputs and rely on powerful hardware to reduce inference latency. Notable examples include \cite{baruch2024polynomial}, \cite{de2024privacy}, and \cite{10946828}. 
For instance, Baruch et al. \cite{baruch2024polynomial} demonstrated inference of an HE-friendly ResNet-50 model using an NVIDIA A100-SXM4-80GB GPU, an AMD EPYC 7763 64-core CPU, and 200GB of RAM.
In parallel, other works focus on improving the accuracy and efficiency of encrypted models by optimizing non-linear function evaluation. Common strategies include hybrid approaches such as combining  CKKS  with TFHE for non-linear activations \cite{294619}, \cite{njungle2025activate} and alternative privacy-preserving techniques like secure multi-party computation \cite{diaa2024fast}. 
While these research directions tackle distinct challenges, research in both directions also become easier with \system.
\system provides a modular and configurable foundation with core HE-friendly CNN components, enabling seamless integration of future advances in throughput optimization and activation function evaluation, while also lowering the barrier for researchers to experiment with new improvements as well as focus their research mainly on aspects that matter to them.

%% file: sections/background.tex
\section{Background}
\label{sec:background}

\subsection{Homomorphic Encryption}
Homomorphic Encryption (HE) is a cryptographic technique that enables computations to be performed directly on encrypted data. 
It reflects the mathematical property where operations in the encrypted (ciphertext) domain yield results that correspond to equivalent operations performed in the unencrypted (plaintext) domain.  
Fully Homomorphic Encryption (FHE), was long regarded as the "holy grail" of cryptography, until 2009 when Craig Gentry demonstrated its feasibility \cite{gentry}. Gentry introduced the concept of \textit{bootstrapping}, a technique used to refresh the noise in a ciphertext through secure re-encryption, effectively allowing for unrestricted computations. This process ensures that the ciphertext maintains sufficient noise tolerance, enabling further computations without compromising its decryptability and security. 

\begin{figure}[http]
    \begin{center}
    \includeimg{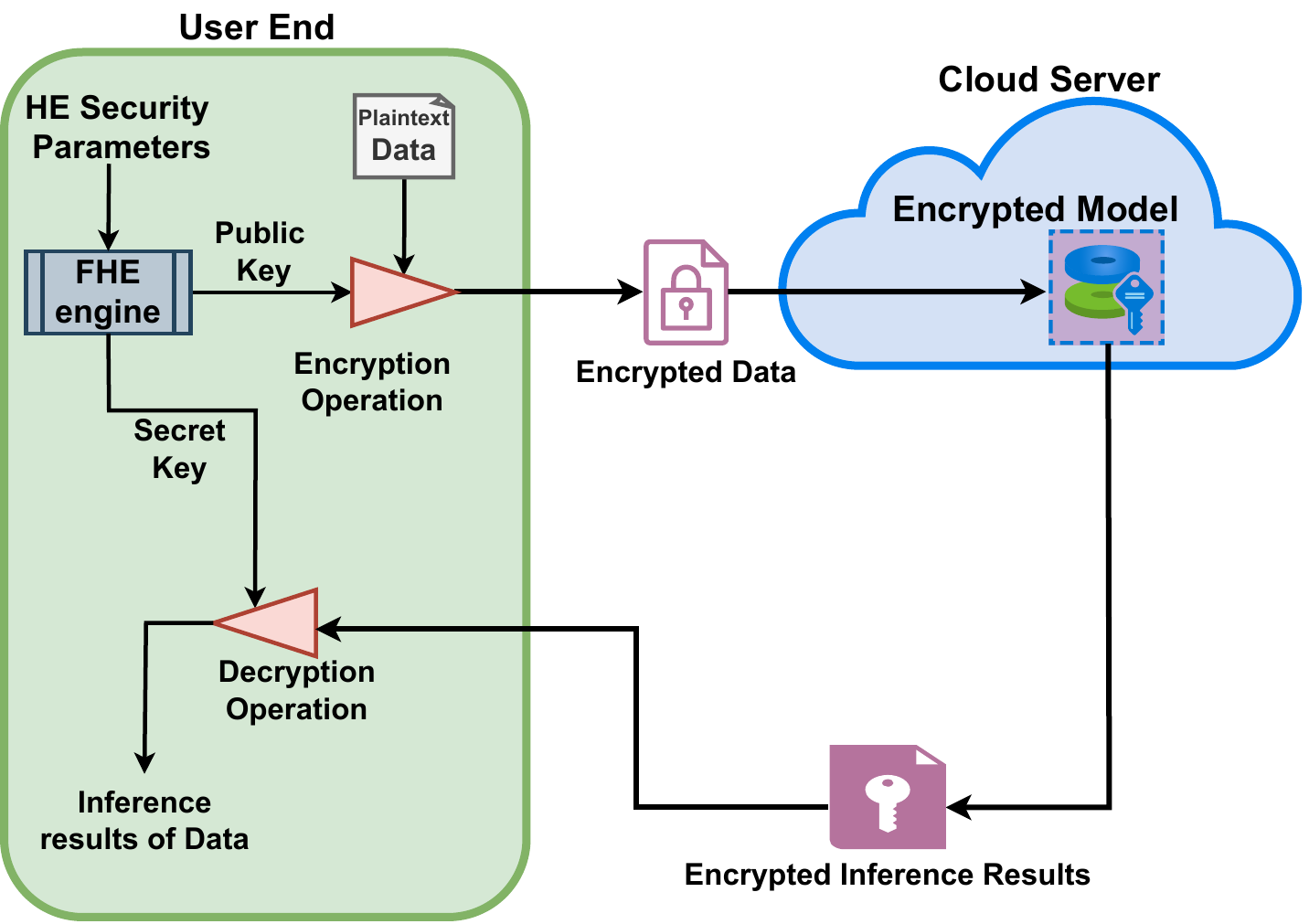}
    \captionsetup{justification=centering}
    \caption{ML inference in a cloud scenario showing a user inferring encrypted data over an encrypted model.}
    \label{fig:fheworks}
    \end{center}
\end{figure}

Figure \ref{fig:fheworks} illustrates an outsourced machine learning inference scenario using HE. 
When a user needs to perform inference on an encrypted model stored in the cloud, the user encrypts their data using their public key and transfers it to the cloud for inference. 
The cloud server then evaluates this data securely over the encrypted model and sends the encrypted results back to the user, who then decrypts the data using their private key. 
The data encrypted and used for inference and decrypted results are only available on the user end, thus preserving the privacy of the inferred data from both the cloud provider and model owner. 

\subsection{Cheon-Kim-Kim-Song (CKKS) Scheme}

Among the various FHE schemes proposed, the CKKS scheme stands out as the most suitable for ML applications as it is the only mainstream scheme supporting approximate arithmetic, uniquely aligned with the requirements of modern ML workflows.
Its security is rooted in the non-probabilistic Ring Learning with Errors (RLWE) problem, an extension of the Learning with Errors (LWE) problem introduced by Regev in 2005~\cite{lwepropsed} and further formalized for polynomial rings by Lyubashevsky et al.~\cite{ringlwe}. 
At its core, the LWE problem involves reconstructing secret information from noisy linear equations generated using random samples. Mathematically, let $\mathbb{Z}$, $\mathbb{R}$, and $\mathbb{C}$ denote the sets of integers, real numbers, and complex numbers, respectively. Given a vector $b \in \mathbb{Z}_q^m$ and a matrix \(A \in \mathbb{Z}_q^{m \times n}\), the standard LWE problem seeks an unknown vector \(s \in \mathbb{Z}_q^n\) such that:
\begin{equation}
    A s + e = b \; \mathrm{mod} \; q,
\end{equation}
where \(e\) is a noise vector sampled from a specific error distribution, and \(q\) is a large prime modulus.

The LWE problem extended to polynomial rings over finite fields, enabling the CKKS scheme to leverage the algebraic structure of polynomials for computational efficiency. This allows support for "slot packing" where multiple plaintext data elements are encoded into the coefficients of a single polynomial. This packing enables the Single Instruction Multiple Data (SIMD) paradigm in CKKS. 
Let \(R = \mathbb{Z}[X] / (X^N + 1)\) represent a ring of polynomials modulo \(X^N + 1\), where \(N\) is a power of 2. Given a message vector \(v \in \mathbb{C}^{N/2}\), CKKS maps \(v\) into the ring \(R\), utilizing at most half of the ring’s coefficients to store data. This encoding ensures that polynomial multiplication in the ciphertext space corresponds to element-wise multiplication in the plaintext space, making the scheme highly efficient for parallel computations.
CKKS defines addition, multiplication, rotation, and bootstrapping operations for the manipulation of encrypted data.

The SIMD paradigm enables parallel data processing by simultaneously applying a single operation across multiple data elements, all encoded within a single ciphertext. 
Homomorphic operations are performed in parallel across all slots, greatly enhancing computational throughput. 
However, due to CKKS's reliance on complex number encoding, at most half of the available slots of a ciphertext can be independently utilized. This limitation arises from the need to maintain conjugate symmetry in encoding to ensure correct decryption. 
Despite this limitation, the SIMD feature boost performance gains through parallel processing of large-scale data, making it practical and scalable for real-world applications.

In this work, we leverage the OpenFHE library's implementation of CKKS \cite{OpenFHE} which is based on an residual number systems (RNS-CKKS) introduced in \cite{cheon2018full} . OpenFHE provides all the CKKS operations outlined in this literature as well as other advanced optimizations such as Fast Rotations \cite{fast_rotations}, further boosting the scheme's operational efficiency within the library. At the same time, OpenFHE development community is the most active HE development community.

%% file: sections/threat_model.tex
\section{Threat Model}

The threat model assumed by \system is consistent with most prior works  leveraging HE for privacy and security guarantees. Specifically, we adopt a \textit{semi-honest} (honest-but-curious) adversary model. Under this model, both participating entities which are typically a client and a server, are assumed to faithfully follow the prescribed protocol without deviation. However, while they do not launch active attacks such as message tampering or denial-of-service, they may attempt to infer additional information about each other’s private data by analyzing the encrypted computation and ciphertext.

In our setting, we assume that the server holds the model weights in plaintext, as is typical in MLaaS deployments where the model is owned or maintained by the service provider. The client encrypts its input locally and transmits only ciphertexts to the server. 
The server then performs inference directly over encrypted inputs using its plaintext model parameters and returns the encrypted prediction to the client. Since only the client possesses the decryption key, only the client can recover the final inference result. 
This setting protects the confidentiality of client data while allowing the server to leverage its proprietary models without modification.

Although \system could be extended to support encrypted model weights thereby also protecting model confidentiality, we do not pursue this setting in the present work.
Our focus is on ensuring strong privacy for client data in the honest-but-curious model, while maintaining practical performance and memory efficiency for encrypted inference.
Furthermore, \system like most HE-based works, does not defend against active adversaries or side-channel attacks, nor does it address availability attacks such as denial-of-service. These types of attacks also remain out of scope for our current work.

%% file: sections/cnnlayers.tex
\section{Secure Neural Network Layers}
\label{sec:cnnlayers}

This section provides an overview of the main CNN layers and operations provided by \system for the development of privacy-preserving HE-friendly models based on CKKS. 
By examining the fundamental components of CNNs, existing literature in integrating them with HE, we provide configurable optimized, and efficient designs and implementations of these layers. 
Our goal is to provide a framework where ML users can easily develop efficient privacy-preserving CNN models.

\subsection{Secure Convolution Layer} 

The convolution operation is a data processing technique used to extract features from input data \cite{namatevs2017deep}. In deep learning, the convolutions play a key role in automatically and adaptively learning spatial hierarchies of features by extracting them from its input data. A Convolution utilizes filters, also known as kernels, which are applied to the input data by sliding them across the spatial dimensions of the layer's input data. At each position, the convolution operation, which is an element-wise multiplication, is performed between the kernel and the corresponding region of the input data, followed by a summation of the results to produce a single output value \cite{Kamath2019}.
Mathematically
let the input tensor be represented by \( X \), with dimensions \( (C, H, W) \), where \( C \) corresponds to the number of input channels, \( H \) is the height, and \( W \) is the width of the input matrix.
The kernel \( K \) has dimensions \( (F, C, k_h, k_w) \), where \( F \) represents the number of output channels, \( C \) is the number of input channels, \( k_h \) is the height of the kernel, and \( k_w \) is the width of the kernel.
The convolution operation can be mathematically expressed as:
{\small
\begin{align}
    Y_{f, h_{\text{out}}, w_{\text{out}}} = 
    \sum_{c=1}^{C} 
    \sum_{i=1}^{k_h} 
    \sum_{j=1}^{k_w} 
    X_{c, h_{\text{out}} + i - 1, w_{\text{out}} + j - 1} 
    K_{f, c, i, j} + b_f
\end{align}
}
\( X_{c, h_{\text{out}} + i - 1, w_{\text{out}} + j - 1} \) is the value of the input tensor at the corresponding input channel, height, and width indices. \( K_{f, c, i, j} \) is the value of the kernel associated with the output channel \( f \), input channel \( c \), and kernel indices \( i \) and \( j \). The term \( b_f \) is the bias added to the result for the output channel \( f \).
\\
The convolution operation produces the output feature map \( Y \), with dimensions \( (F, H_{\text{out}}, W_{\text{out}}) \) where the height and width of the output tensor are defined by:
\begin{align}
    H_{\text{out}} &= H - k_h + 1, \\
    W_{\text{out}} &= W - k_w + 1.
\end{align}

Performing convolution operations naively within FHE infrastructure poses significant challenges due to the substantial storage and computational resources required, since every value in the output tensor will require a distinct computation. 
 To address this, Juvekar et al. introduced an optimized method known as Vector Encoding in \cite{gazelle}. This technique exploits SIMD (Single Instruction, Multiple Data) capabilities to perform convolution operations more efficiently, reducing memory overhead and improving overall performance. Vector Encoding has since become a standard in second-generation FHE-based CNNs, as seen in works such as \cite{kim2023optimized}, \cite{lee2022privacy}, and \cite{rovida_cnn}.
A common limitation in literature is that the encoding is often optimized for the specific model being evaluated. 
In contrast, our work generalizes this technique and implemented a configurable convolution layer that supports arbitrary input and output channels  and also  incorporates different padding and striding techniques. 
 Our approach therefore provide an adaptable and configurable HE-friendly convolution layer similar to what we see in the broader field of ML thus, applicable to all classes and architectures of CNNs.

\begin{figure}[http]
    \begin{center}
    \includeimg{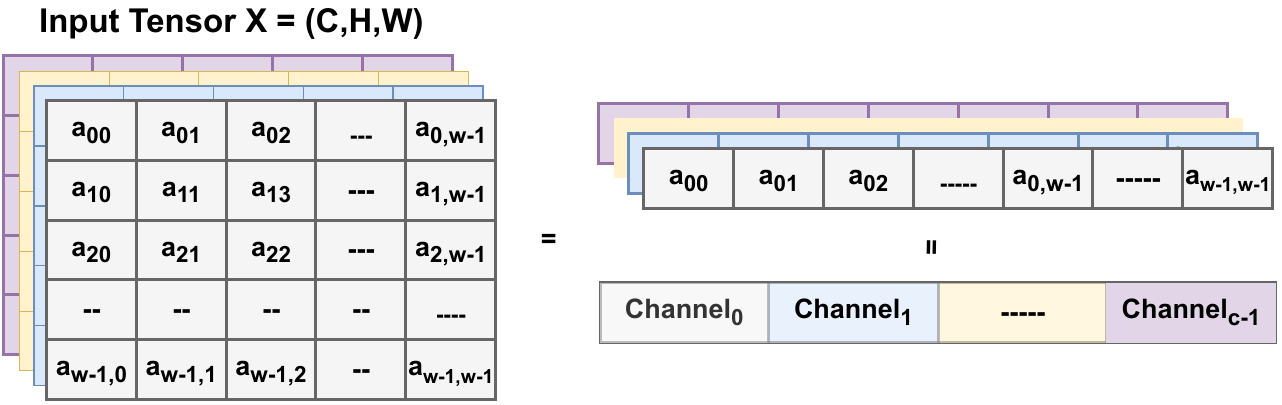}
    \captionsetup{justification=centering}
    \caption{Flattening of an Input tensor into a SIMD Vector to be encrypted and used as an input of \system }
    \label{fig:simd_flattened}
    \end{center}
\end{figure}

The convolution layer receives a flattened input tensor in form of a single SIMD ciphertext. 
In the first convolution layer of most CNNs, this is the flattened encrypted input. 
In subsequent layers within the network, it is the output of the previous layer. 
It also receives, the input dimension (that is input width), kernel size, input channels, and output channels configurations.
Following this, \(k^2 - 1\) rotations of the input ciphertext are generated, where \(k\) denotes the kernel width. These rotations align all the elements of the input tensor that correspond to the first convolution operation as the first elements of the rotated ciphertexts. Each rotated ciphertext is then multiplied by a corresponding vector, which contains a repeated kernel value of the corresponding rotated ciphertext. The original ciphertext is also multiplied by a vector that contains the repeated value of the first element of the kernel matrix. The results of these multiplications are summed to produce the convolution output for the entire input tensor. 
Figure \ref{fig:simd_flattened} shows how an input tensor \( X \) with dimensions \( (C, H, W) \) is been flattened and aligned channel by channel as a SIMD Vector to be used as input to the first secure convolution layer.  

\begin{figure}[http]
    \begin{center}
    \includeimg{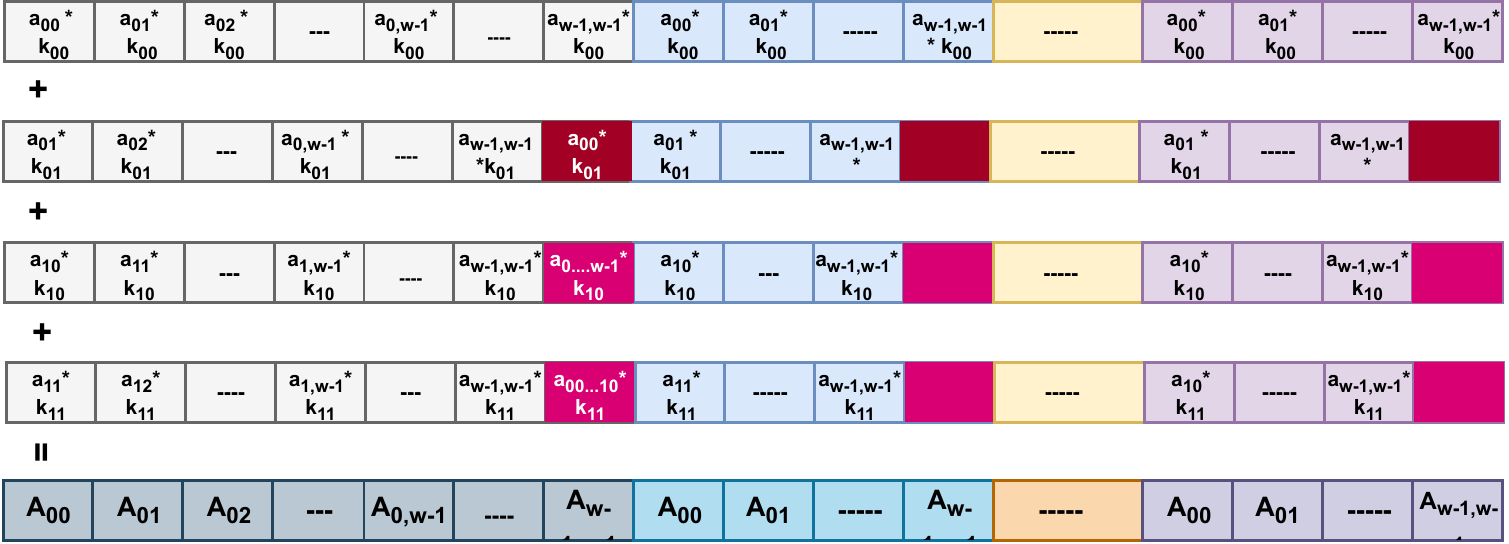}
    \captionsetup{justification=centering}
    \caption{Rotated SIMD ciphertexts multiplied with their equivalent kernel vectors and summed to produce the \( A \) ciphertext.}
    \label{fig:conv_a}
    \end{center}
\end{figure}

In the second step, each rotated ciphertext is multiplied by the equivalent repeated kernel element vectors, and the resulting ciphertexts are summed to produce a ciphertext that contains the convolution result of the input ciphertext, denoted as \( A \), as shown in Figure \ref{fig:conv_a}.
Just like the input channels are flattened and aligned channel by channel, the repeated kernel value vectors of input channels are also flattened and aligned in a similar fashion.  
This ensure that kernel data for all input channels matches the appropriate data in the input tensor. This alignment allows the convolution operation to be applied for all elements across all input channels simultaneously, further improving our convolution layer's efficiency. 
This Stage requires a single multiplication for every kernel element and $k^2-1$ additions to sum the results of the different multiplications into a new ciphertext.

After the summation of the ciphertexts, for every output channel,  we construct \( C \) ciphertexts from \( A \), where each ciphertext corresponds to the convolution operation on an input channel.
We optimize this step to reuse the single rotation key of \( W^2 \). This is done \(C\) times where the ciphertext of $i-1$ is the input for generating the  corresponding ciphertext of $i$ where $i < C$. 
We then add all the $ C $ constructed ciphertext to give a resulting ciphertext that contains the correct values of the convolution layer for that output channel.  We then remove all the unwanted information from this intermediate ciphertext by multiplying it with a mask containing $w^2$ of $1s$ and $C-1$ $0s$, producing a ciphertext that shall be referred to as $A^*$.

\begin{figure}[http]
\vspace{-0.1in}
    \begin{center}
    \includeimg{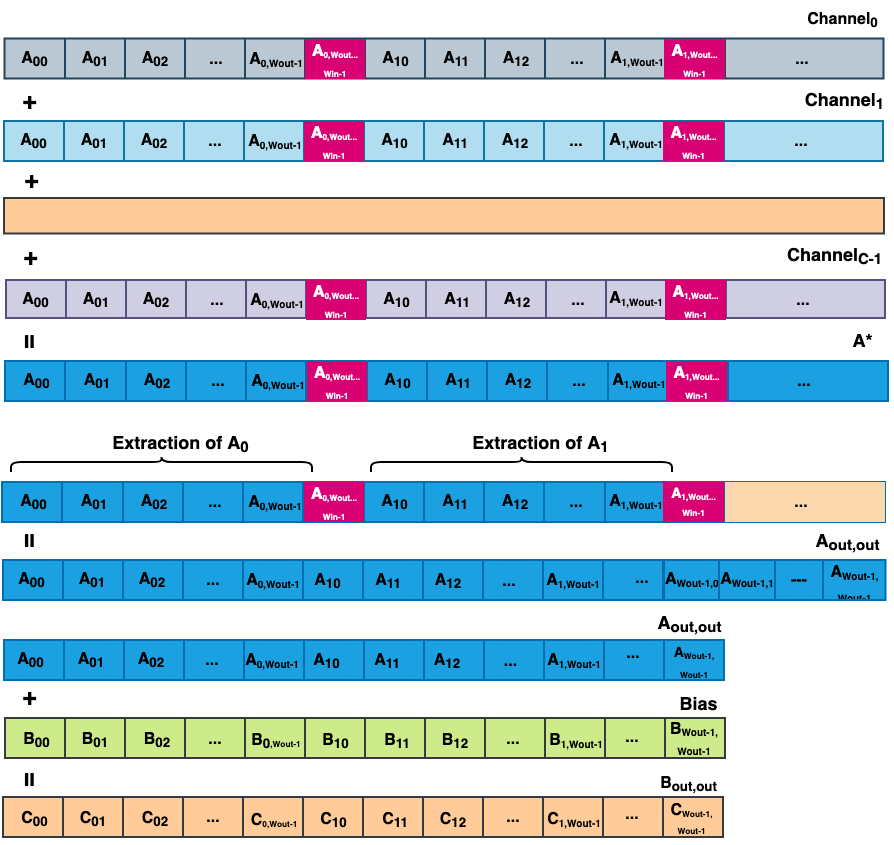}
    \captionsetup{justification=centering}
    \caption{Secure convolution using SIMD showing the extraction process and addition of bias}
    \label{fig:secure_con}
    \end{center}
\end{figure}

To obtain the result, \( W_{\text{out}} \), additional rotations of the resultant ciphertext are performed to extract all elements of \( A_{\text{out}, \text{out}} \). 
To further optimize this process, we also reuse a single rotation key of $ W_{\text{out}}$ to extract all the information across all output widths. 
This process is repeated for all output channels. For the remaining \( F-1 \) output channels, negative rotations of \(W^2_{out}\) are applied to their ciphertexts, and the results are added into a single ciphertex to construct the final feature map ciphertext \( Y \). 
An equivalent bias vector is then added to form the final ciphertext. 
As illustrated in Figure \ref{fig:secure_con}, $A$ generates $C-1$ rotated ciphertexts of all input channels. This ciphertext is summed into $A^*$. Then, the extraction process is conducted to produce $A_{out,out}$, and the Bias vector $B_{out, out}$ is added to produce the final results $C_{out, out}$.
This approach to the convolution layer ensures efficient computation of the convolution operation under HE constraints with the  efficient re-use of resources while preserving the capabilities and correctness of all operations. Our approach ensures that every convolution requires exactly (\( K + F \)) re-usable rotation keys.

\subsection{Secure Advanced Convolution Layer}
The convolution layers perform the essential task of feature extraction from input data by applying the kernels over the input tensor. With additional parameters such as padding and striding, the convolution layer controls the spatial dimensions of the output.
Padding is a technique used to preserve the spatial dimensions of the input tensor. By adding extra rows and columns of values (typically zeros) around the input tensor, padding ensures that the convolution operation can extract more information from edge regions to contribute to the feature map and can prevent premature spatial downsampling, leading to a loss of information.
On the other hand, Striding refers to the step size at which the kernel is moved across the input tensor to perform the convolution operation. A stride of $1$ moves the kernel one unit at a time, resulting in a densely computed feature map. Larger stride values, such as $2$ or more, lead to a downsampled output by skipping intermediate computations, reducing the dimensions of the feature map. 

The choice of striding and padding values is crucial in CNNs as they directly affect the performance and feature preservation in models.
The convolution layer with stride value $S$ and padding value $P$ produces the output feature map \( Y \), with dimensions \( (F, H_{\text{out}}, W_{\text{out}}) \). The height and width of $Y$ are given by Equations \ref{eq:convhi} and \ref{eq:convwi}, respectively.
\begin{align}
    H_{\text{out}} &= ((H + 2P  - k_h)/S)+1, \label{eq:convhi} \\
    W_{\text{out}} &= ((W + 2P - k_w)/S)  + 1. \label{eq:convwi}
\end{align}

\subsubsection{Secure Advanced  Convolution With Padding}
Padding generally modifies the dimensions of the input tensor for the convolution operation by adding additional rows and columns of padded values around the input matrix for all input channels. The new input tensor, denoted as \( X_p \), has dimensions \( (C, H_p, W_p) \), where the padded height and width are calculated as shown in Equation \ref{eq:advhi} and \ref{eq:advwi} respectively:
\begin{align}
H_p &= H + 2P,  \label{eq:advhi} \\
W_p &= W + 2P.  \label{eq:advwi}
\end{align}

Adding padding to the encrypted input tensor introduces an additional computation and memory overhead to the secure convolution layer. To optimally implement padding, we leverage the flattened input tensor's sequential nature, understanding that each row's first element is adjacent to the last element of the previous row in the flattened representation. 
This property allows us to efficiently add horizontal and vertical padding using a single rotation for every two sequential rows of the same input channel. 
Specifically, we rotate the ciphertext by \( W \) to align rows and then append \( 2P \) padded  \( 0 \) values in the appropriate positions for all $H$. These rotations introduce padded $0$ values by default to the ciphertext.

After this initial step, additional \( W_p + P \) values of \( 0 \) are appended at the beginning and end of each input channel in the flattened ciphertext to complete the padding process. We rotate for each input channel using $W^2$ and then use one rotation with a key of \( W_p + P \) to introduce $0s$, which correspond to the first row and first element in the padded ciphertext. We automatically gain the  \( W_p + P \) $0$s values for the last channel at the end of the ciphertext since the SIMD vector encodes $0s$ in all unused slots. Since this process is repeated independently for all input channels  \( C \), it requires only $W$ rotations per channel.
This optimized padding approach efficiently integrates the padding values while minimizing the number of rotations and rotation keys required by using features such as the structural properties of the flattened input tensor in the SIMD ciphertext. The padded ciphertext $X_p$ is then input into the convolution layer, with the new width and height calculated as the padded width and height, ensuring that the resulting feature map accurately reflects the effects of the padding. 
Our process requires \( C+3 \) rotation keys, significantly reducing the memory overhead that can be introduced by traditional padding in HE.

\subsubsection{Secure Advanced Convolution With Striding}
\label{subsec:secure_convolution_with_striding} 

Striding is a technique used to reduce the feature map's spatial dimensions by controlling the kernel's step size as it moves over the input. The striding value, denoted as \( S \), determines the number of units the kernel shifts horizontally and vertically during the convolution process. When \( S > 1 \), the convolution operation skips \( S-1 \) units both horizontally and vertically. This reduces the number of computed output values, effectively down-sampling the feature map.
In the SIMD-based convolution process described earlier, the convolution operation for all elements in the input tensor is calculated at once; thus, striding introduces an additional processing overhead in selecting the right output values. Initially, all convolution-resulting values for every output channel are calculated with a single multiplication operation and additions resulting in the intermediate ciphertext \( A^* \), which contains the convolution results for a striding value of \( S = 1 \). 
We implement two approaches for striding offering different levels of flexibility, computational overhead and memory overheads. 

In the first approach we minimizes the noise level in the encrypted ciphertext making it easier to evaluate deeper models with fewer bootstrappings.  Here, we perform an extraction process to isolate only the convolution values corresponding to the desired striding value \( S > 1 \).
The striding process involves skipping values both vertically and horizontally in the ciphertext.
Starting with the vertical striding process, for each vertical index \( i \), if \( i \mod S = 0 \) and \( i < H_{\text{out}} \), we extract the required convolution values by rotating \( A^* \) by \( W \times i \), where \( W \) is the width of the input tensor. This operation aligns the convolution values for the \( i \)-th row into an intermediate ciphertext \( T_i \).
In the Horizontal Striding process, for each horizontal index \( j \), we extract the first \( j \)-th values in \( T_i \). By applying additional rotations of \( S \times j \), where \( S \times j < W_{\text{out}} \), we extract all right convolution values on the ciphertext. The extracted values are then merged to form the horizontally strided convolution values for the \( i \)-th row of the output feature map.
After extracting the horizontal values for each row, we apply negative rotations of $W_{out}$ to align the needed values and add them to produce the final output ciphertext for that corresponding channel output \( A_{\text{out}, \text{out}} \). 
This process is repeated for all output channels, ensuring that the resulting feature map accurately reflects the specified striding value.
This approach to striding uses $W_{out} + 2$ rotation keys, which is still minimal compare overhead for this process while maintaining the integrity and functionality of the convolution process with striding within the HE infrastructure, even for large and complex input tensors and kernels.

In the second approach, we implement an optimized striding method that significantly reduces the number of ciphertext rotations by strategically using multiplicative masks. 
The use of multiplicative masks drastically use up the ciphertext noise budget compared to the first approach but lead to better runtime performancve and reduced memory overhead since it requires fewer rotation keys.
The method begins by generating a binary mask of size $W^2$. This mask contains ones in the positions corresponding to stride-aligned elements and zeros elsewhere. By multiplying the input ciphertext with this mask, we effectively eliminate all non-stride positions from the ciphertext, reducing unnecessary data early in the process.
Following this initial pruning, we perform a sequence of $\log_2(W_{\text{out}})$ ciphertext rotations. After each rotation, we apply an additional multiplicative mask to remove redundant or overlapping values. These masks are constructed to preserve only the valid components introduced by each rotation. The resulting ciphertexts are then aggregated to gradually form a compact representation of the strided output.
Finally, a set of $W_{\text{out}}$ row-specific masks is applied in a loop, each isolating a single row in the output grid. These masked ciphertexts are accumulated to create the final compacted ciphertext that contains the correctly downsampled result. This final stage ensures that all retained values are aligned into contiguous slots in the ciphertext.
Overall, this approach requires $\log_2(W_{\text{out}}) + W_{\text{out}}$ ciphertext multiplications and rotations, and uses only $\log_2(W_{\text{out}}) + 1$ unique rotation keys. 

To further optimize the second striding approach, we introduce a configurable pipeline that performs striding across multiple output channels simultaneously, without incurring additional memory overhead. 
Specifically, instead of applying convolution and striding channel by channel, we first compute the convolution for a group of $g$ output channels, where $g$ is set equal to the number of input channels of that layer. 
The resulting intermediate outputs are then concatenated, and striding is applied jointly across all $g$ output channels.
This design yields several advantages in the HE setting. First, it eliminates the need for generating rotation keys that would otherwise be used only once, thereby reducing the memory overhead. Second, it allows the the ciphertext ring dimension and number of slots to be constrained to the exact number of elements required within the CNN model.

\subsection{Special Secure Convolution Layer}
In many modern neural network architectures such as ResNet and VGG architectures, the convolution layers are often configured to take a padding of \(1\), a striding of \(1\), and a kernel with a size of \(3 \times 3\). These configurations ensure that the input tensor and output tensor dimensions remain the same. 
These configurations of the convolution layer give rise to a special case where we can leverage the structures of the input and output to use very few rotations by avoiding the padding of the input tensor but still carrying out the convolution operation with padding. 
This approach has been used by works like Rovida et al.  \cite{rovida_cnn} to efficiently optimize the performance of their ResNet-20 model. 
Understanding that there is no change in input and output tensor shapes, we avoid the expensive padding of the input tensor and the extraction step of the convolution layer, creating a special convolution case. 
Instead, we perform special rotation operations on the input ciphertext to create the required rotated ciphertexts for corresponding repeated kernel vector multiplications pre-determined for this special case. 
These rotations of ciphertext $c$ are performed to yield $r_i$ corresponding ciphertexts as demonstrated in the Algorithm \ref{alg:precompute}. 
$r_i$ is multiplied with equivalent kernel vectors of the repeated kernel values as shown in Figure \ref{fig:conv_a}.

\begin{algorithm}[H]
\caption{Rotated Ciphertexts for Convolution}
\label{alg:precompute}
\begin{algorithmic}[1]
\REQUIRE Ciphertext $c$, input width $w$
\ENSURE Rotated ciphertexts $r_0, \ldots, r_8$
\STATE $r_4 \leftarrow c$
\STATE $r_1 \leftarrow \text{Rot}_{-w}(r_4)$
\STATE $r_7 \leftarrow \text{Rot}_{w}(r_4)$
\STATE $r_3 \leftarrow \text{Rot}_{-1}(r_4)$
\STATE $r_5 \leftarrow \text{Rot}_{+1}(r_4)$
\STATE $r_0 \leftarrow \text{Rot}_{-1}(r_1)$
\STATE $r_2 \leftarrow \text{Rot}_{+1}(r_1)$
\STATE $r_6 \leftarrow \text{Rot}_{-1}(r_7)$
\STATE $r_8 \leftarrow \text{Rot}_{+1}(r_7)$
\RETURN $r_0, r_1, \dots, r_8$
\end{algorithmic}
\end{algorithm}

To introduce the corrections needed in the output of this layer, we optimized the algorithm to dynamically introduce $0$s in the vectors that contain the repeated kernel values required to generate the right results regardless of the shape of the input tensor or number of channels. Algorithm \ref{alg:mask_gen} shows how we generate binary masks of $1$s and $0s$ needed to multiply with the kernel vectors for all input channels. This mask introduces $0s$ in the kernel vectors, which directly maps to all unwanted operations in the SIMD convolution operation. We then use  Algorithm \ref{alg:all_mask} to generate the exact mask for every kernel vector of repeated values. This mask is then multiplied with the repeated value kernel vectors of $W^2$ elements for each channel to generate the optimized kernel vectors passed to the convolution layer. After multiplying the special rotations shown above with the optimized kernel vectors, for every output channel, we use just a single rotation key of $W^2$ to align and sum the convolution results of all input channels. Lastly, we add the bias vector creating a ciphertext whose results are equal to that of the convolution layer with all the configurations but without explicitly performing the padding and extraction steps of the convolution layer. 

\begin{algorithm}
\caption{Generation of a mask repeated for all input channels in a function called build\_mask}
\label{alg:mask_gen}
\begin{algorithmic}[1]
\STATE \textbf{Input:} $sp$ (starting padding), $ep$ (ending padding), $w$ (window width), $m$ ($W^2$), $C$ (input channels)
\STATE \textbf{Output:} $\textit{tmask}$ (all channels mask)

\STATE \textbf{Initialization:}  
Initialize an empty list $mask$

\FOR{$i = 0$ \TO $sp - 1$}
    \STATE Append $0.0$ to $mask$
\ENDFOR

\WHILE{size of $mask < (m - ep)$}
    \FOR{$j = 0$ \TO $w - 1$}
        \STATE Append $1.0$ to $mask$
    \ENDFOR
    \STATE Append $0.0$ to $mask$
\ENDWHILE

\WHILE{size of $mask > m$}
    \STATE Remove last element from $mask$
\ENDWHILE

\WHILE{size of $mask < m$}
    \STATE Append $0.0$ to $mask$
\ENDWHILE

\FOR{$i = 0$ \TO $ep - 1$}
    \STATE Set $mask[m - i - 1] \gets 0.0$
\ENDFOR

\STATE Initialize an empty list $\textit{tmask}$

\FOR{$i = 0$ \TO $t - 1$}
    \STATE Append $mask$ to $\textit{tmask}$
\ENDFOR

\STATE \textbf{return} $\textit{tmask}$
\end{algorithmic}
\end{algorithm}

\begin{algorithm}
\caption{Generate $9$ Masks for Special Convolution Layer}
\label{alg:all_mask}
\begin{algorithmic}[1]
\STATE \textbf{Input:} $m$ ($W^2$), $C$
\STATE \textbf{Output:} $\textit{all\_masks}$ (set of 9 masks)

\STATE Initialize $\textit{all\_masks}$ as an empty list

\STATE Append the following masks to $\textit{all\_masks}$:
\STATE $\quad \textit{build\_mask}(W + 1, 0, W - 1, m, C)$
\STATE $\quad \textit{build\_mask}(W, 0, m, m, C)$
\STATE $\quad \textit{build\_mask}(W, 0, W - 1, m, C)$
\STATE $\quad \textit{build\_mask}(1, 0, W - 1, m, C)$
\STATE $\quad \textit{build\_mask}(0, 0, m, m, C)$
\STATE $\quad \textit{build\_mask}(0, 1, W - 1, m, C)$
\STATE $\quad \textit{build\_mask}(1, W - 1, W - 1, m, C)$
\STATE $\quad \textit{build\_mask}(0, W, m, m, C)$
\STATE $\quad \textit{build\_mask}(0, W + 1, W - 1, m, C)$

\STATE \textbf{return} $\textit{all\_masks}$
\end{algorithmic}
\end{algorithm}

\subsection{Secure Average Pooling Layer}

The pooling layer is primarily used for dimensionality reduction \cite{gholamalinezhad2020pooling}. It is conceptually similar to convolution with striding but differs in that its filters are uniform. Two widely used pooling layers in CNNs are \textit{Maximum Pooling} and \textit{Average Pooling}. Maximum Pooling selects the maximum value from each region that the pooling kernel covers. At the same time, Average Pooling computes the average of all values within the region covered by the pooling kernel \cite{PASSRICHA20195}.
Pooling layers generally use a stride of at least $2$, allowing the filter to skip certain input regions, effectively reducing the data size in computation for subsequent network layers.
In our context with CKKS, \textit{Maximum Pooling} is computationally expensive as it involves evaluating a non-linear greater-than function. In contrast, \textit{Average Pooling} is well-suited for encrypted computations as it can be performed using a single multiplication and addition, as shown in Equation \ref{eq:pooling}. 
$k$ is the pooling kernel width, $k^2$ is the size of the pooling kernel.
\begin{align}
\text{Output}(i, j) = \frac{1}{k^2} \sum_{x=0}^{k-1} \sum_{y=0}^{k-1} \text{Input}(i+x, j+y), \label{eq:pooling}
\end{align}

\begin{figure}[http]
    \begin{center}
    \includegraphics[width=\linewidth]{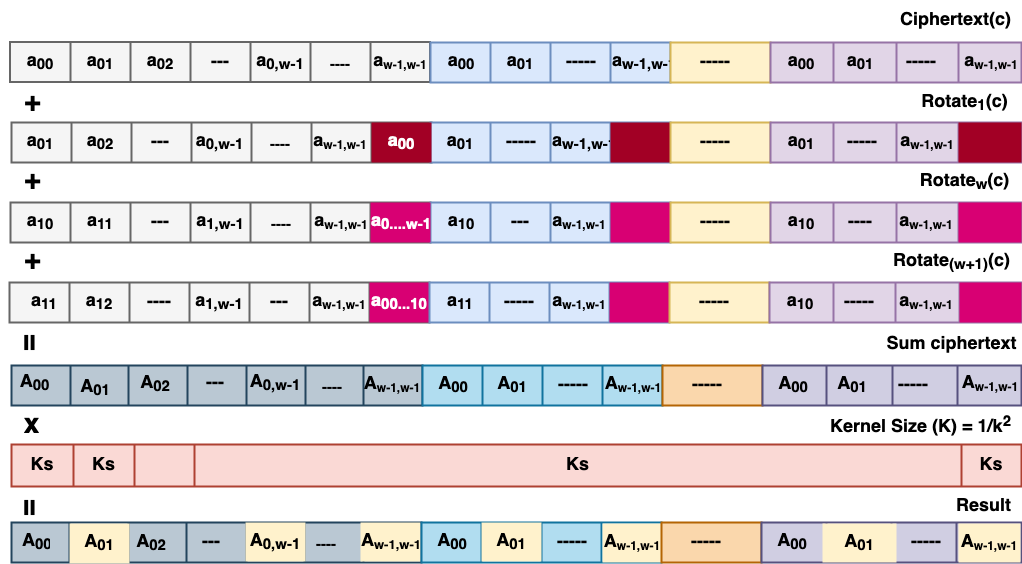}
    \captionsetup{justification=centering}
    \caption{Secure Average Pooling with kernel scaling value of \(\frac{1}{k^2}\). The resulting ciphertext is passed to a striding module.}
    \label{fig:secure_pooling}
    \end{center}
\end{figure}

To implement average pooling in the encrypted domain, \system uses the same algorithm as in the convolution layer to generate the different rotations required from the encrypted input. 
We then sum the rotated ciphertexts and multiply the resulting ciphertext by the precomputed value of $\frac{1}{k^2}$. This pre-computation value of $\frac{1}{k^2}$ efficiently introduces a division of $k^2$, ensuring that the average pooling values for the entire input tensor are calculated. 
Figure \ref{fig:secure_pooling} demonstrates our approach for average pooling where $k = 2$; thus, the value of $\frac{1}{k^2} = 0.25$ is pre-computed. We then follow the same algorithm as in the convolution layer with striding. We multiply our  $A^*$ ciphertext with a values of $\frac{1}{k^2}$ to introduce the required scaling. 

We also implement a special case of Average Pooling called Global Average Pooling, which is very popular in the ResNet architectures. In this unique type of pooling, instead of reducing the spatial dimensions by taking the average over small regions, the pooling layer operates over the entire spatial dimension of a feature map, reducing it to a single value per feature per channel. In \system, we implement this algorithm by summing all values of the input ciphertext within $W^2$. We do this for all channels and merge the results into a single ciphertext multiplied by the pre-computed repeated vector of $\frac{1}{W^2}$. The resulting ciphertext of this operation is equal to the output of the Global Average Pooling layer.

Furthermore, in cases where the kernel size is less than or equal to the width of the input channel, we handle it as a special case. We implement this by multiplying the input ciphertext with a pre-computed repeated vector of $\frac{1}{k^2}$. We then repeatedly use one rotation key to extract each channel's first value. We do this by rotating the ciphertext by $W^2$ and merging the resulting vectors. 
The resulting ciphertext of this operation is also equal to the output of the pooling layer.

\subsection{Secure Fully Connected Layer}

A fully connected layer in CNNs maps all input data to an output space by learning a set of weights that define the relationships between every input and output neuron \cite{intro_cnn}. In this layer, each input neuron is connected to every output neuron, enabling the network to capture complex global patterns.
Mathematically, given an input vector \( x \) of size \( n \), weights \( W \) of size \( m \times n \), biases \( b \) of size \( m \), and output vector \( y \) of size \( m \), it's computation is expressed as shown in Equation \ref{eq:fc}:
\begin{align}
    y_k = \sum_{i=1}^{n} W_{ki} x_i + b_k, \quad \text{for } k = 1, 2, \dots, m. \label{eq:fc}
\end{align}

In \system, the use of SIMD packing within the ciphertext makes the implementation of this layer highly efficient, requiring only a single homomorphic multiplication and \( n \) additions for each output neuron. For each output neuron, a SIMD multiplication is performed as \( W_i \times x_i \), where \( W \) represents the weights vector associated with the output neuron, \( x \) is the input neurons vector, and \( i \) denotes the index of the input neuron. This is followed by \( n \) homomorphic additions to calculate the final value for the output neuron. This process is repeated \( m \) times to compute all the output neurons in the layer.
To consolidate the results, \( m \) rotations are used to combine the values of the output neurons into a single ciphertext representing the outputs of the fully connected layer. These \(m \) rotations require an equal number of rotation keys for the respective indices, which can result in significant memory usage for layers with many output neurons. 

To reduce the number of keys required by this layer, we introduce a parameterized merging approach for managing the output ciphertext. This involves configuring the fully connected layer with an additional parameter that specifies the number of available conservative rotation keys within the system. By doing so, \system optimizes the rotation operations required to compute the final outputs of the layer while reducing the memory overhead from traditional approaches.

\subsection{Activation Function (Secure ReLU)}
Non-linear layers are also known as activation functions, and they play a crucial role in CNNs by allowing them to capture and represent complex, non-linear patterns in the data \cite{DUBEY202292}.
Among various activation functions used in CNNs, the Rectified Linear Unit (ReLU) is one of the most widely adopted due to its simplicity and effectiveness \cite{he2018relu}.
It applies a threshold operation and set all negative values to $0$s, defined as shown in Equation \ref{eq:relu}. 
This non-linearity accelerates convergence and also helps reduce the vanishing gradient problem in training. 
It's approximation using Chebyshev polynomials has also been widely adopted for use with HE-friendly CNNs. 
\begin{align}
\label{eq:relu}
    ReLU(x) = max(0, x)
\end{align}

Chebyshev polynomials are a family of orthogonal polynomials defined over the interval \([-1, 1]\). They are particularly valuable for approximating complex non-linear functions efficiently and accurately thus have been widely adopted for evaluating non-linear operations in HE \cite{gil2012non}. 
Denoted by \(T_d(x)\), these polynomials are defined recursively as follows:
\begin{align}
    T_0(x) &= 1, \\
    T_1(x) &= x, \\
    T_d(x) &= 2x T_{d-1}(x) - T_{d-2}(x), \quad \text{for } d \geq 2.
\end{align}

One of the notable properties of Chebyshev polynomials is the distribution of their roots. A polynomial of degree \(D\) has exactly \(d\) roots, all lying within the interval \([-1, 1]\). These roots are obtained as defined by Equation \ref{eq:cheroots}:
\begin{align}
    x_k = \cos\left(\frac{k\pi}{d}\right), \quad \text{where } 0 \leq k < d. \label{eq:cheroots}
\end{align}

This structured arrangement of roots allows Chebyshev polynomials to achieve optimal uniform approximation for continuous functions over the interval \([-1, 1]\). 
To approximate a non-linear function \(f(x)\) over the interval \([-1, 1]\), it can be expressed as a truncated series of Chebyshev polynomials:
\begin{align}
    f(x) \approx \sum_{d=0}^{D} c_d T_d(x),
\end{align}
where \(c_d\) are the coefficients derived from \(f(x)\), and \(D\) is the degree of the Chebyshev polynomial. 

To apply Chebyshev approximations correctly, it is essential to normalize the input data to \([-1, 1]\). 
We introduced a scaling factor \(\beta\) that adjusts the input vector accordingly. 
The scaling process is done by multiplying the input vector by \(\frac{1}{\beta}\), as shown in Equation \ref{eq:scale_relu}.
Our design allows the users to set the scaling factor $\beta$, which should be chosen based on an analysis of their dataset and model. 
After scaling, the adjusted ciphertext is passed to the Chebyshev approximation function. The results are then multiplied by $\beta$, as shown in Algorithm~\ref{alg:secure_relu}.
\begin{align}
\label{eq:scale_relu}
    x_{\text{s}} = \frac{1}{\beta} \cdot x
\end{align}

\begin{algorithm}
\caption{Secure ReLU Using Homomorphic Encryption}
\label{alg:secure_relu}
\begin{algorithmic}[1]
\STATE \textbf{Input:} $x$, $\beta$, $n$
\STATE \textbf{Output:} $y$

\STATE \textbf{Initialization:}  
 $D \gets 59$, $v \gets x$

\IF{$\beta > 1$}
    \STATE $p \gets \text{GenerateScaleMask}(\beta, n)$
    \STATE $v \gets \text{EvalMult}(x, p)$
\ENDIF

\STATE Define $f(z): f(z) \gets 0 \text{ if } z < 0; \; f(z) \gets \beta \cdot z \text{ otherwise}$

\STATE $y \gets \text{EvalChebyshevFunction}(f(z), v, D)$

\STATE \textbf{return} $y$
\end{algorithmic}
\end{algorithm}

In this algorithm, $n$ denotes the size of the input vector, which varies based on the previous layer of the network, and $D$ represents the degree of the Chebyshev polynomial used for function approximation. 
Through experimentation with different values of $D$ on the ResNet-20 and LeNet-5 architectures, we determined that $D = 59$ provides on average, a computationally efficient balance for CNNs while maintaining high accuracy. This setting is used for all the models evaluated in this work and set as the default value in \system but just like other parameters, it is also configurable.

%% file: sections/fhelayers.tex
\subsection{FHEON'S Support Functions}

To support the development of complete HE-friendly privacy-preserving CNN models, \system provides numerous support functions. 
In this paper, we have described the most useful set of these functions, but a complete list  shall be made available as part of \system's documentation upon publication.

\subsubsection{Context Generation}
The context generation layer is responsible for initializing the HE cryptographic environment required for secure computation. This layer abstracts the complexity of CKKS parameter tuning, making the framework accessible to both novice users with little to no background in cryptography as well as advanced users who require fine-grained control over their encryption settings.
At its core, the layer accepts essential user-defined parameters such as the polynomial ring dimension, number of slots, and scaling factor. 
These parameters are internally mapped to the appropriate settings in the OpenFHE library to construct a valid HE cryptographic context.

Once the context is initialized, the layer automatically generates the associated cryptographic keys.
The generated keys can be serialized and stored locally in a user-specified location for ease of use and reproducibility. Users are advised to keep their private key confidential, as it is required to decrypt the output of the encrypted inference process.
In addition to the required inputs, users may optionally specify advanced parameters such as the number of slots, the scaling mode, scaling factor,   multiplicative depth, and the rescaling strategy used during homomorphic operations. If any of these parameters are omitted, \system falls back on a set of well-balanced default values that aim to provide an optimal trade-off between computational performance and accuracy.

\subsubsection{Encryption Layer}
The encryption layer is responsible for securely encoding and encrypting user data based on the previously generated FHE context and public key. This layer acts as the bridge between plaintext data and its encrypted counterparts, ensuring data confidentiality throughout the computation process. It accepts an input as a matrix, flattens it into a vector. It then encodes the vector into a CKKS plaintext polynomial and encrypts it using the public key of the user. It returns this encrypted ciphertext to the user, who can then inference it through a HE-friendly model without revealing any information about the input to the server or model  during the computations.

\subsubsection{Weights Loading}
Model weights in \system are processed per layer. 
\system provides two primary mechanisms for integrating model weights, designed to balance development-time efficiency with runtime flexibility. 
Model weights are ingested in CSV format, making \system model's independent of the underlying training methodology. 
Exporting model weights from frameworks such as PyTorch or TensorFlow for \system is straightforward, as no additional preprocessing or reformatting is required.
The first mechanism follows a preprocessing strategy, where all model weights are loaded into the system prior to inference. To support this, we provide utility functions that accept CSV files along with the corresponding kernel shapes and channels configurations, automatically reshaping and formatting the data to match the model’s internal structure. 
This enables seamless weight initialization and is well-suited for scenarios where performance is critical as well as continuous inferring scenarios. The trade-off is increased memory usage, since all weights must be loaded into memory. 

The second mechanism enables dynamic weight loading, where weights are formatted and injected into the model at runtime. 
This approach is more memory-efficient and offers greater flexibility, making it suitable for resource-constrained systems or applications requiring on-the-fly model adaptability. 
However, it introduces a computational trade-off, since model weights must be continuously processed, loaded and offloaded from memory during the inference process.

\subsubsection{Evaluation Keys Management}
The Evaluation Keys Management helper functions in \system are responsible for orchestrating the generation, serialization, deserialization, and efficient usage of evaluation keys required during encrypted inference. 
Specifically, the evaluation keys includes rotation keys, which enable slot-wise data movement in ciphertexts and bootstrapping keys, which are required to refresh ciphertexts and extend computation depth during model evaluation. 
We also offer helper functions associated with all the CNN layers discussed in \ref{sec:cnnlayers} that receives layer configurations and  automatically determine and generate the unique  rotation indices needed for that layer.  
The equivalent rotation keys for the indices can then be generated. 

Just like weights loading, keys can be loaded into a model at a preprocessing phase or using a dynamic approach. 
In the preprocessing approach, \system analyzes the entire CNN architecture to identify the complete set of unique rotation indices and bootstrapping configurations required for all layers. 
It then generates a unified set of evaluation keys, ensuring no redundant keys are created for the application. 
This method is ideal for performance-critical scenarios as all necessary keys are preloaded into memory.
It is the easiest way for users to load keys and outsourced key management within their model. 
However, this approach introduce higher memory usage.

In the dynamic approach, model keys are changed during runtime. We provide helper functions to support on-demand evaluation key loading and removal. 
This mode is particularly useful for advanced users who understand the structure and execution flow of their model and can strategically manage its keys into blocks that reduce memory overhead. 
It can be very beneficial in inferring on large models. 
By supporting both preprocessing and dynamic key strategies, the Evaluation Keys Management layer provides flexibility to balance performance, memory efficiency, and user control over the HE-friendly privacy-preserving models build on \system. 

\subsubsection{Bootstrapping Layer}
Bootstrapping is an important concept in HE as it allows for arbitrary computations. We provide a helper function that maps the CKKS bootstrapping implementation in OpenFHE to the user. This layer can be easily called at any stage within the network on the ciphertext. It then Bootstrap the ciphertext an allow more room for computations. 

%% file: sections/experiment.tex
\section{Evaluation and Results}
\label{sec:experiment}

All experiments in this study were performed on a consumer-grade CPU (AMD Ryzen 9 5900X 12-core processor with 64 GB of RAM). This processor was selected due to its widespread availability, affordability, and accessibility. 
Unlike many existing works which rely on high-cost hardware, \system is a significant step toward making HE-friendly privacy-preserving ML feasible for commonly used and everyday systems.
We developed \system using OpenFHE v1.2.4 (released March 2025) and have also tested it with v1.2.3, v1.3.0, and v1.3.1. 
\system will be actively maintained, with continuous support for new features, compatibility updates to align with future OpenFHE releases, as well as ongoing contributions to serve the user community.

\subsection{HE Security Parameters}

For all experiments, we picked the CKKS parameters that enable efficient computations across all HE-friendly CNNs models.
Balancing the cryptographic parameters is essential to achieve both efficiency and security. To ensure fair and consistent comparisons, we adopted a  uniform set of configurations, with the exception of a smaller polynomial degree and fewer slots for the LeNet-5 to better match its architecture. Specifically, we used a multiplicative depth of 25, a first modulus size of 50, a rescaling factor of 46, 4 digits for key switching,  and the \texttt{flexibleauto} rescaling strategy. Table~\ref{tab:fhe_params} summarizes the remaining parameters.

\begin{table}[ht]
    \small
    \centering
    \caption{FHE parameter sets for encrypted inference.}
    \label{tab:fhe_params}
    \begin{tabularx}{\linewidth}{|l|l|X|}
        \hline
        \textbf{Parameter} & \textbf{LeNet-5} & \textbf{ResNets + VGGs} \\
        \hline
        Polynomial Degree         & 16,384 & 32,768 \\
        \hline
        Number of SIMD Slots      & 8,192  & 16,384 \\
        \hline
    \end{tabularx}
\end{table}

\subsection{Models Evaluated }
We validate our work across a wide range of CNN architectures, covering the most commonly used CNN classes. These include networks from the LeNet, VGG, and ResNet classes. Each class represents a distinct level of complexity and is generally applied to different problem domains, making them ideal for validating \system's effectiveness and efficiency.

LeNets are characterized by their relatively simple and shallow architectures.  They were designed for grayscale images and are commonly used for tasks involving simpler datasets, such as the MNIST dataset \cite{lenets}. LeNet-5 is the most widely used variant in CNN literature; thus we developed a LeNet-5 model and validated it on the grayscale MNIST dataset. 

The VGG family was introduced in 2015 and has also been extensively used in CNN literature due to their effectiveness in large-scale image tasks \cite{simonyan2015deepconvolutionalnetworkslargescale}. These networks employ deeper architectures with uniform convolutional layers. VGG-11 and VGG-16 are among the most commonly implemented models from this class of models. 
While VGG-11 has $11$ layers, VGG-16 is a deeper architecture with similar characteristics, having 16 layers. We implemented these two architectures from these classes and used the CIFAR-10 dataset to validate them.

The ResNet family is the most recent of the CNN classes \cite{he2015deepresiduallearningimage}. It introduces the concept of residual connections to improve the training of deep networks by addressing the vanishing gradient problem.
In previous HE-friendly CNN works, ResNet-20 has been pervasively studied due to its balance of model depth and performance. 
For this work, we implement ResNet-20 and ResNet-34 models. We evaluate the ResNet-20 model using the CIFAR-10 dataset, while the ResNet-34 model is validated on the more complex CIFAR-100 dataset.

By selecting these representative architectures and datasets across various levels of complexity, we ensure that our methods are validated on a broad spectrum of tasks.
Our evaluation of different datasets, ranging from the simple grayscale MNIST dataset to the very complex CIFAR-100, also allows us to evaluate the robustness and efficiency of \system's models across different datasets. 
Table \ref{tab:arch_com_2} and \ref{tab:arch_com} in the Appendix show all our CNNs architectures, along with their  configurations used in this study. 
Also, Figure \ref{fig:ResNet Basic Block} showcases the basic block structure used by the ResNets models.  It contains two convolution layers with an additional third convolution layer only used in down-sampling blocks (blocks that use striding to reduce the dimensionality of the input tensor). We have two and three of these additional convolution layers in the architectures of ResNet-20 and ResNet-34, respectively \cite{resnet34}.

For this work, we implemented our HE-friendly models using the building blocks provided by \system. For the LeNet-5 model, all model weights and evaluation keys were preloaded during an initial preprocessing phase. 
In contrast, for the deeper ResNet and VGG architectures, we adopted the dynamic loading strategies to balance memory efficiency with runtime performance. 
Specifically, these models were divided into blocks based on their downsampling stages, and for each block, the corresponding evaluation keys were serialized and loaded on demand while keys from the previous layers are cleared from memory. 
We further tailored our implementations to the architectural characteristics of each model. 
For example, the ResNet models utilized the specialized secure convolution layers while LeNet-5 employed the generic secure convolution layers.
To address ciphertext noise growth, we incorporated Bootstrapping operations at strategically chosen points in the networks. In particular, Bootstrapping was applied prior to every ReLU activation (except for the first two), before the second pooling layers, and prior to the GlobalPooling layer. The placement of these Bootstrapping operations was determined for every model based on the observed noise levels in the ciphertexts during inference.

All our models were developed and trained using Python and PyTorch. 
During training, we utilized batch normalization after each convolution for all models except for the LeNet-5. 
Batch normalization is used as it improves the training of the networks and keeps the range of data within the network relatively small and stable \cite{ioffe2015batchnormalizationacceleratingdeep}. 
After training, the learned batch normalization weights were folded back into their respective convolutional kernels weights before exporting them \cite{EdouardYvinec-2022}.

\subsection{Results}
Table \ref{tab:accuracy_comparison} illustrates the accuracy of different CNN models evaluated under plaintext and encrypted settings. Notably, our approach shows consistent performance with minimal degradation in accuracy across the models. The table also shows the number of images inference in the encrypted domain for each model.
The encrypted LeNet-5 and ResNet-20 models are within \(\substack{+\\-}\) \(1\%\) of the corresponding plaintext accuracies of these models. 
Table \ref{tab:memory_usage} presents the memory utilization for all our models. \system, demonstrates balanced memory usage and inference speed across all models.  
The reported values are computed by selecting a random image from the validation set and averaging the memory usage and inference time over ten independent inference runs.
These values were observed to be close range across the full validation set.

\begin{table}[http]
    \small
    \centering
    \caption{Accuracy of models in the Plaintext and FHE settings.}
    \label{tab:accuracy_comparison}
    \begin{tabularx}{\linewidth}{|X|X|X|X|}
        \hline
        \textbf{Architecture} & \textbf{Plaintext(\%)} & \textbf{Encrypted(\%)}  & \textbf{Images}\\
        \hline
        LeNet-5  & 98.8 & 98.5 & 1000 \\
        \hline
        ResNet-20 & 92.8 & 92.2 & 560 \\
        \hline
         ResNet-34   & 76.5 & 74.4 & 210 \\
        \hline
         VGG-11 & 87.7 & 86.1 & 331 \\
        \hline
        VGG-16   & 88.3 & 85.8 & 120\\
        \hline
    \end{tabularx}
\end{table}

\begin{table}[http]
    \small
    \centering
    \caption{Memory Usage and Inference Time}
    \label{tab:memory_usage}
    \begin{tabularx}{\linewidth}{|X|X|X|}
        \hline
        \textbf{Architecture} & \textbf{Memory(GB)} & \textbf{Inference Time (s)} \\
        \hline
        LeNet-5  & 4.2 & 13 \\
        \hline
        ResNet-20 & 20.4 & 403  \\
        \hline
         ResNet-34   & 25.6 & 1594 \\
        \hline
        VGG-11 & 39.6 & 1647 \\
        \hline
        VGG-16   & 42.3 & 2232 \\
        \hline
    \end{tabularx}
\end{table}

To contextualize our findings within existing research, we benchmarked the ResNet-20 model built on \system against state-of-the-art models on the CIFAR-10 dataset. A summary of this comparison is provided in Table~\ref{tab:related_works_comparison}.

\begin{table}[http]
    \centering
    \small
    \caption{Comparison with Related Works using ResNet-20}
    \label{tab:related_works_comparison}
    \begin{tabularx}{\linewidth}{|l|X|X|X|}
        \hline
        \textbf{Proposal} & \textbf{Acc. (\%)} & \textbf{Mem. (GB)} & \textbf{Lat. (s)} \\
        \hline
        Woo et al. \cite{lee2022privacy}             & 90.67 & 512+ & 10602 \\
         \hline
        Lee et al. \cite{lee2022low}  & 91.31 & 512+ & 2271 \\
         \hline
        
        Kim et al. \cite{kim2023optimized} & 92.04 & 100  & 255 \\
         \hline
        Rovida et al. \cite{rovida_cnn}   & 91.53 & 21.8 & 410 \\
         \hline
        \textbf{\system}              & \textbf{92.2} & \textbf{20.4} & \textbf{403} \\
        \hline
    \end{tabularx}
    \vspace{0.5em}
    
    \footnotesize\textit{Processor Summary:} \\
    Lee et al. : Threadripper PRO 3995WX  \\
   Woo et al. : Dual Intel Xeon Platinum 8280 (112 cores), \\
    Kim et al. : EPYC 7402P, \\
    Rovida et al. / \system: AMD Ryzen 9 5900X 12-core processor
\end{table}
\system’s ResNet-20 model achieved the lowest memory overhead reported in the encrypted domain while also delivering competitive accuracy and runtime. For a fair comparison, we benchmarked our system against the state-of-the-art work of Rovida et al., who reported an inference time of 260 seconds on Apple’s M1 Pro processor. Running their open-source implementation on the same AMD Ryzen hardware as our experiments, we measured an inference time of 410 seconds with a memory footprint of 21.8 GB per image. 
\system's ResNet-20 model achieved an average inference time of 403 seconds with a reduced memory footprint of 20.4 GB.
Rovida et al. employ a highly specialized design, embedding preprocessed hard-coded weights directly into the model and adopting an optimized vector-encoding strategy with carefully selected rotation indices. 
While these optimizations are effective, they are also tightly tailored to fixed scenarios and thus less generalizable across diverse architectures. By contrast, \system offers flexibility, configurability, and generality in developing HE-friendly CNNs.  
While dynamically processing weights and allocating resources at runtime,  the generalized model we build on \system is still extremely efficient.
In addition, \system offers flexibility in importing weights and supports multiple evaluation-key reuse strategies. 

Very few HE CNN works report results for VGGs and ResNet-34 models. 
Those that do are generally not performance centric thus, their results are generally underwhelming and incomplete. 
For instance Junghyun et al. \cite{lee2023precise} introduced a min-max approach for ReLU and MaxPool approximation, achieving 88.7\%, 88.2\%, 89.1\% and 91.1\% on ResNet-20, ResNet-34, VGG-11, and VGG-16 respectively. Runtime are only reported for the ResNet-20 (2,892s), and ResNet-34 (4,764s) with no memory profiling for the rest of the models.

%% file: sections/conclusion.tex
\section{Conclusion}
\label{sec:conclusion}
In this work, we introduced \system, a configurable and highly optimized framework for privacy-preserving CNN models  development using HE. 
We used \system to develop various well-established CNN models, including LeNet-5, VGG-11, VGG-16, and ResNet-20, and ResNet-34.
Notably, our results show that even complex CNN architectures, like VGG-16 and ResNet-34, requires  consumer-grade hardware with less than 64 GB of memory and achieve an inference time within practical constraints.
Furthermore, comparisons with prior works highlight substantial reductions in resource consumption as well as comparable performance with highly optimized state-of-the-art works, underscoring the efficiency and scalability of \system.
These findings establish \system as a robust solution for the development of HE-friendly CNN models, offering a balance of high accuracy and resource efficiency. 
By offering configurable CNN layers designed on CKKS, \system bridges the gap between theoretical advancements in HE and practical use in ML. Its compatibility with consumer-grade hardware strengthens its viability for widespread adoption. 

While additional features such as encrypted weight processing are important, future research will also focus on optimizing \system to improve computational efficiency and enhance its practicality in resource-constrained environments. 
Improving throughput and further reducing accuracy degradation through better approximation mechanisms as well as noise management mechanisms are also interesting works that should be incorporated into \system. 
Moreover, extending support to neural network layers not yet implemented  would also expand the framework’s applicability to other domains such as privacy-preserving transformers. 
Finally, enabling privacy-preserving training through encrypted backpropagation and stochastic gradient descent  also presents a compelling avenue for research, bringing \system closer to supporting end-to-end privacy-preserving model development under HE constraints.
\system is openly available at 
\url{https://github.com/stamcenter/fheon}, with a comprehensively
documentation provided for users at 
\url{https://fheon.pqcsecure.org}.

%% file: sections/appendix/architectures.tex
\section{Model Architectures}

\begin{table}[http]
\centering
\small
\renewcommand{\arraystretch}{1.2}
\caption{Architectural Comparison of ResNet-20 and ResNet-34. Downsampling convolutions with a $1\times 1$ kernel, a stride of $2$, and a padding of $1$ are omitted from this table. Acronyms: Residual Blocks (RBs)}
\label{tab:arch_com_2}
\resizebox{\textwidth}{!}{%
\begin{tabular*}{\textwidth}{|p{0.07\textwidth}|p{0.1\textwidth}|C{0.0625\textwidth}|C{0.065\textwidth}|C{0.08\textwidth}|}
\cline{1-5}
\textbf{Network}  & \centering \textbf{Layer} & \makecell{\textbf{Input}\\\textbf{Channels}} & \makecell{\textbf{Output}\\\textbf{Channels}} & \makecell{\textbf{Kernel Size,}\\\textbf{Stride,}\\\textbf{Padding}}  \\    
\cline{1-5}
\multirow{6}{*}{\textbf{ResNet-20}}        
    & Conv + ReLU & 3 & 16  & $3 \times 3$, 1, 1 \\ \cline{2-5}
    & 3 RBs & 16 & 16 & $3 \times 3$ , 1, 1 \\ \cline{2-5}
    & 3 RBs & 16 & 32 & $3 \times 3$ , 1 , 1 \\ \cline{2-5}
    & 3 RBs & 32 & 64 & $3 \times 3$ , 1 , 1 \\ \cline{2-5}
    & GlobalAvgPool       & 64  & 64  & Global       , 1 , 0 \\ \cline{2-5}
    & FC            & 64  & 10  & N/A \\ \cline{1-5}
\multirow{7}{*}{\textbf{ResNet-34}}        
    & Conv + ReLU & 3 & 64  & $7 \times 7$ , 2 , 3 \\ \cline{2-5}
    & 3 RBs & 64 & 64 & $3 \times 3$ , 1 , 1 \\ \cline{2-5}
    & 4 RBs & 64 & 128 & $3 \times 3$ , 1 , 1 \\ \cline{2-5}
    & 6 RBs & 128 & 256 & $3 \times 3$ , 1 , 1 \\ \cline{2-5}
    & 3 RBs & 256 & 512 & $3 \times 3$ , 1 , 1 \\ \cline{2-5}
    & GlobalAvgPool       & 512 & 512  & Global , 1 ,0 \\ \cline{2-5}
    & FC            & 512  & 10  & N/A  \\ \cline{1-5}
\end{tabular*}
}
\end{table}

\begin{table}[http]
\centering
\vspace{0.1in}
\small
\renewcommand{\arraystretch}{1.2}
\caption{Architectural of LeNet-5, VGG-11, and VGG-16. All convolution (conv) and fully connected (FC) layers, except for the last layer of the network, are followed by the ReLU activation function.}

\label{tab:arch_com}
\begin{tabular*}{\textwidth}{|p{0.06\textwidth}|p{0.09\textwidth}|C{0.0625\textwidth}|C{0.0625\textwidth}|C{0.09\textwidth}|}
\cline{1-5}
\textbf{Network}  & \centering \textbf{Layer} & \makecell{\textbf{Input}\\\textbf{Channels}} & \makecell{\textbf{Output}\\\textbf{Channels}} & \makecell{\textbf{Kernel Size,}\\\textbf{Stride,}\\\textbf{Padding}}  \\
\cline{1-5}
\multirow{7}{*}{\textbf{LeNet-5}}        
    & Conv & 1  & 6   & $5 \times 5$ , 1 , 0 \\ \cline{2-5}
    & AvgPool & 6   & 6  & $2 \times 2$ , 2 , 0 \\ \cline{2-5}
    & Conv & 6   & 16  & $5 \times 5$ , 1 , 0 \\ \cline{2-5}
    & AvgPool & 16  & 16  & $2 \times 2$ , 2 , 0 \\ \cline{2-5}
    & FC   & 256 & 120  & N/A \\ \cline{2-5}
    & FC   & 120 & 84  & N/A \\ \cline{2-5}
    & FC   & 84 & 10  & N/A \\  \cline{1-5}
\multirow{16}{*}{\textbf{VGG-11}}          
    & Conv & 3   & 16  & $3 \times 3$ , 1 , 1 \\ \cline{2-5}
    & AvgPool & 16   & 16  & $2 \times 2$ , 2 , 0 \\ \cline{2-5}
    & Conv & 16  & 32 & $3 \times 3$ , 1 , 1 \\ \cline{2-5}
    & AvgPool & 32   & 32  & $2 \times 2$ , 2 , 0 \\ \cline{2-5}
    & Conv & 32  & 64 & $3 \times 3$ , 1 , 1 \\ \cline{2-5}
    & Conv & 64  & 64 & $3 \times 3$ , 1 , 1 \\ \cline{2-5}
    & AvgPool & 64   & 64  & $2 \times 2$ , 2 , 0 \\ \cline{2-5}
    & Conv & 64  & 128 & $3 \times 3$ , 1 , 1 \\ \cline{2-5}
    & Conv & 128 & 128 & $3 \times 3$ , 1 , 1 \\ \cline{2-5}
    & AvgPool & 128   & 128  & $2 \times 2$ , 2 , 0 \\ \cline{2-5}
    & Conv & 128  & 128 & $3 \times 3$ , 1 , 1 \\ \cline{2-5}
    & Conv & 128 & 128 & $3 \times 3$ , 1 , 1 \\ \cline{2-5}
    & AvgPool & 128   & 128  & $2 \times 2$ , 2 , 0 \\ \cline{2-5}
    & FC  & 1024 & 1024  & N/A \\ \cline{2-5}
    & FC  & 1024 & 1024  & N/A \\ \cline{2-5}
    & FC  & 1024 & 10  & N/A \\ \cline{1-5}
\multirow{21}{*}{\textbf{VGG-16}}          
    & Conv & 3   & 16  & $3 \times 3$ , 1 , 1 \\ \cline{2-5}
    & Conv & 16   & 16  & $3 \times 3$ , 1 , 1 \\ \cline{2-5}
    & AvgPool & 16   & 16  & $2 \times 2$ , 2 , 0 \\ \cline{2-5}
    & Conv & 16  & 32 & $3 \times 3$ , 1 , 1 \\ \cline{2-5}
    & Conv & 32  & 32 & $3 \times 3$ , 1 , 1 \\ \cline{2-5}
    & AvgPool & 32   & 32  & $2 \times 2$ , 2 , 0 \\ \cline{2-5}
    & Conv & 32  & 64 & $3 \times 3$ , 1 , 1 \\ \cline{2-5}
    & Conv & 64  & 64 & $3 \times 3$ , 1 , 1 \\ \cline{2-5}
    & Conv & 64  & 64 & $3 \times 3$ , 1 , 1 \\ \cline{2-5}
    & AvgPool & 64   & 64  & $2 \times 2$ , 2 , 0 \\ \cline{2-5}
    & Conv & 64  & 128 & $3 \times 3$ , 1 , 1 \\ \cline{2-5}
    & Conv & 128  & 128 & $3 \times 3$ , 1 , 1 \\ \cline{2-5}
    & Conv & 128  & 128 & $3 \times 3$ , 1 , 1 \\ \cline{2-5}
    & AvgPool & 128   & 128  & $2 \times 2$ , 2 , 0 \\ \cline{2-5}
    & Conv & 128  & 128 & $3 \times 3$ , 1 , 1 \\ \cline{2-5}
    & Conv & 128  & 128 & $3 \times 3$ , 1 , 1 \\ \cline{2-5}
    & Conv & 128  & 128 & $3 \times 3$ , 1 , 1 \\ \cline{2-5}
    & AvgPool & 128   & 128  & $2 \times 2$ , 2 , 0 \\ \cline{2-5}
    & FC    & 1024 & 1024 & N/A \\ \cline{2-5}
    & FC   & 1024 & 1024 & N/A \\ \cline{2-5}
    & FC   & 1024 & 10  & N/A \\ \cline{1-5}
\end{tabular*}
\end{table}

\begin{figure}[http]
    \centering
    \includegraphics[width=0.7\linewidth]{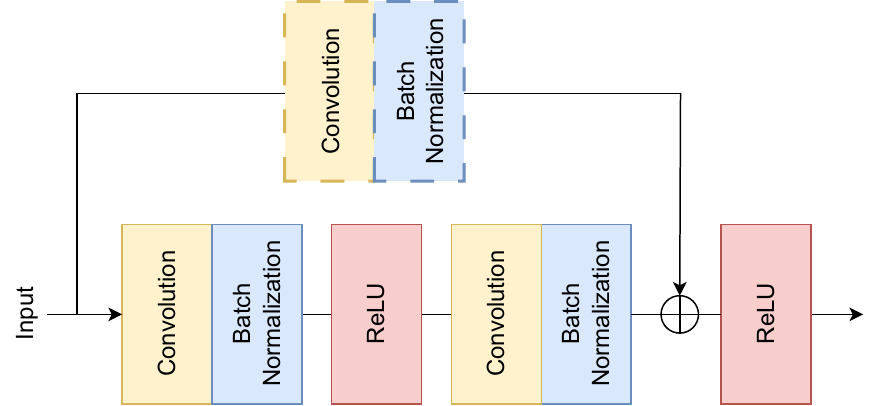}
    \caption{Basic Block Structure of ResNet. Dashed blocks signify that these operations only occur after downsampling.}
    \label{fig:ResNet Basic Block}
    \vspace{0.1in}
\end{figure}